\documentstyle[aps,preprint]{revtex}
\begin{document}

\baselineskip=20pt
\thispagestyle{empty}

\vspace{20mm}

\begin{center}

{\Large Thermal Relaxation Time in Chemically Non-equilibrated
Quark- Gluon Plasma}

\vskip 0.3cm

 Xiao-Fei Zhang $^{a,b}$, \ \  Wei-Qin  Chao$^{a,b}$

\footnotesize{\sl $^a$CCAST (World Laboratory), P. O. Box 8730, Beijing 100080,
People's Republic of China

$^b$Institute of High Energy Physics, Academia Sinica, Beijing 100039,
People's Republic of China  }


\begin{minipage}{100mm}
\vspace{3cm}
\centerline{\bf Abstract}
\vspace{5mm}

The definition of  thermal relaxation time is extended to 
chemically non-equilibrated quark-gluon  plasma 
and the chemical non-equilibrated thermal relaxation times
for partons are  calculated using the non-equilibrium Debye mass 
as the infrared  regulator. The dependence of the thermal relaxation 
time on the fugacity is  given and the influence of the chemical 
non-equilibration is discussed.
We find that 
there are  threshold fugacities   $\lambda_g^*$ and 
$\lambda_q^*$ for gluons and quarks.
for $\lambda_g<\lambda_g^*$($\lambda_q<\lambda_q^*$),
$\tau_g^{NEQ}/\tau_g^{EQ}$ ($\tau_q^{NEQ}/\tau_q^{EQ}$) 
decreases strongly with increasing 
fugacity, while for $\lambda_g>\lambda_g^*$($\lambda_q>\lambda_q^*$),
the ratios are almost 1.
It is shown   that there is also the  two-stage equilibration 
in a chemically non-equilibrated plasma. We also discussed  the effect
of using the non-equilibrium  Deby mass as the infrared cutoff.

\end{minipage}
\end{center}

\vspace{20mm}
PACS number(s): 12.38.Mh, 24.85.+p, 52.25.Dg

\newpage

{\large \bf{I.  INTRODUCTION}}

\medskip

One of the main objectives of the future experimental programs at the 
BNL Relativistic Heavy Ion Collider (RHIC) and the CERN Large Hadron
Collider (LHC) is the production of Quark-Gluon Plasma (QGP) through 
heavy ion collisions[1]. 
The parton gas produced in heavy ion collisions is not immediately in thermal 
and chemical equilibrium. The parton gas relaxes to the
equilibrium state through secondary interactions.   
 Because the preequilibrium
phase may influence the QGP signals, it is essential to
discuss nonequilibrium properties of QGP and to investigate whether 
QGP produced in RHIC and LHC could reach equilibrium.
  Quantum transport theory[2] in principle can be used to
describe the properties of nonequilibrium QGP and 
there are some discussions based on it[3]. However, its applicability
to heavy ion collisions is still far from being realistic.
The   Boltzmann equation  in the relaxation time approximation
is often used to  discuss  the transport coefficients
 and the evolution of the QGP [4-5].
In this approximation method the collision terms are determined by the 
corresponding thermal relaxation time
which measures the time scale of the nonequilibrium
system approaching  equilibrium and can be estimated from the 
parton  interaction 
rates[6-8]. The  equilibration of QGP  considering  of parton
  expansion effect is discussed recently [5]
However, the thermal relaxation time  has been considered 
so far mostly for a chemically equilibrated plasma. 
As  discussed by many models[9-11],
in the early stage of ultra-relativistic heavy ion collisions,
the parton gas is dominated by gluons and  far
from chemical equilibrium. First, the system is
evoluting to   thermal equilibrium mainly through elastic scattering,
which leads to the local isotropy in momentum distribution. 
The system evolute towards  chemical equilibrium later
and sometimes the plasma could never reach chemical equilibrium
during its life time. 
In this  paper we calculate the thermal relaxation time in a 
chemically non-equilibrated  plasma and discuss how the 
chemical nonequilibration  influences the parton thermal relaxation.

The thermal relaxation time in QGP diverges strongly at small momentum 
transfer in the naive perturbation theory. This difficulty could be overcome
by the  Bratten-Yuan method[12] which
 proposed that  
the soft contribution $k<k^*$ and the hard part 
$k>k^*$ could be calculated separately
by introducing a separation scale $k^*$ for  the momentum transfer.
The soft part is treated
using the  resumed propagators and vertices proposed by Braaten and
Pisarski[13], whereas the bare Green functions are sufficient for the hard
one. Assuming $gT<k^*<T$, the final result is independent  of the
separation $k^*$.  There is another much simpler and widely applied approximation
method based on the naive perturbation theory, 
using bare propagators with the Debye mass as infrared regulator[14].
It was  demonstrated that this  method also worked well 
for calculating the thermal relaxation time[6].
We will use this approximation approach, and  
 the nonequilibrium Debye mass as the  infrared regulator
in this paper. 

This paper is organized as follows, in Section II we first define 
the interaction rates in a chemically non-equilibrated plasma.
Then, in Section III, the thermal relaxation time for quarks 
and gluons are calculated using the definition of section II.
Finally, in section  IV, we give some numerical results and
discussions.

\vspace{2cm}
{\large {\bf
 II.  DEFINITION OF THE INTERACTION RATES IN A CHEMICALLY 
NON-EQUILIBRATED GAS} }

In this section we will extend  Weldon's discussion[17] and 
define the interaction rates for 
a chemically non-equilibrated gas. 
There are two kinds of interaction rates. One is the ordinary 
interaction rate
defined as  the damping rate for partons,
which is usually  called interaction rate. 
The other  is the so called transport interaction rate  
obtained from the ordinary one  by 
introducing a transport weight containing the scattering angle in 
the center of mass system.  Because large
angle scattering  is the most efficient mechanism for the dissipative
momentum transfer in the case of a plasma with long range interaction,
hence the mean free path, as well as the thermal
relaxation time, may rather be defined by the inverse of the transport 
interaction rate[15]. Let us first define the ordinary interaction
rate for a chemically non-equilibrated plasma from the Bolzmann 
equation[16],   
 \begin{eqnarray}
v_1\cdot\partial_1f_1(p_1)=&&-{\nu\over 2p_1}\int{d^3 p_2\over (2\pi)^3 2p_2}
\int{d^3 p_3\over (2\pi)^3 2p_3} \int{d^3 p_4\over (2\pi)^3 2p_4}\nonumber\\
&&[f_1f_2(1\pm f_3)(1\pm f_4)-
(1\pm f_1)(1\pm f_2)f_3f_4]\nonumber\\
&&(2\pi)^4\delta^4(P_1+P_2-P_3-P_4)|M_{12\rightarrow 34}|^2,
\end{eqnarray}
where $P_i=(p_i,{\vec p_i})$, $v^{\mu}_i=p^{\mu}/ p_i$.
$\partial_i^\mu=\partial/\partial(x_i)_\mu$.
$f_i$ is the distribution function $f(p_i)$
for partons.
$\pm $ are used for bosons (gluons ) and fermions(quarks and 
anti-quarks).
$|M_{12\rightarrow 34}|^2$ is the squared matrix 
element for corresponding  parton scattering process and  
 is summed over final states and averaged over initial states.
The spin and color factor $\nu$ is 16 for gluons and 6 $N_f$
with $N_f$  flavors of quarks or anti-quarks.

To obtain the definition of the interaction rates[17],
 we assume that $f_1$ is away from  equilibrium and
$f_i, i=2,3,4, $ are in thermal equilibrium but not 
in chemical equilibrium, which  can be 
expressed as  the following form,
 \begin{equation}
f^e_i=\lambda_i{1\over e^{\beta p_i}\mp\lambda_i},\ \ \ \ i=2,3,4,
\end{equation}
where $\lambda_i$ is the fugacity of the corresponding partons.
It measures how far the system is from the chemical 
equilibrium.
Using the relation for the distribution function Eq.(2),
\begin{equation}
1{\pm}f_i^e=\lambda_i^{-1}e^{\beta p_i}f_i^e,
\end{equation}
we obtain from the Boltzmann equation,
\begin{equation}
v_1\cdot \partial_1f_1(p_1)=-\Gamma (p_1)(f_1-f_1^e),
\end{equation}
 \begin{eqnarray}
\Gamma(p_1)=&&{\nu\over 2p_1}\int{d^3 p_2\over (2\pi)^3 2p_2}
\int{d^3 p_3\over (2\pi)^3 2p_3} \int{d^3 p_4\over (2\pi)^3 2p_4}
{f_2^e(1\pm f_3^e)(1\pm f_4^e)\over (1\pm f_1^e)}\nonumber\\
&&(2\pi)^4\delta^4(P_1+P_2-P_3-P_4)|M_{12\rightarrow 34}|^2.
\end{eqnarray}
The above equation is the definition of the interaction rates in a 
chemically non-equilibrated plasma. By 
 putting  $\lambda_i=1$,
Eq.(5) is just the interaction rates for 
a chemically equilibrated plasma[17].

We will assume $f^e_1=f^e_3$,$f^e_2=f^e_4$
in the following discussion. 
This simplification holds as long as 
$k=|{\vec p_1}-{\vec p_3}|=|{\vec p_2}-{\vec p_4}|$
is not too large or $-t $ is not of the order of $s$,
with $s, t, u$ being  the usual Mandelstam variables[8,16].
Now the transport interaction rates  and the thermal 
relaxation time can be defined from Eq.(5) after introducing 
a transport weight $sin\theta^2/2$, 
 \begin{eqnarray}
{1\over\tau}=\Gamma^{trans}
=&&{\nu\over 2p_1}\int{d^3 p_2\over (2\pi)^3 2p_2}
\int{d^3 p_3\over (2\pi)^3 2p_3} \int{d^3 p_4\over (2\pi)^3 2p_4}
f_2^e(1\pm f_4^e)\nonumber\\
&&(2\pi)^4\delta^4(P_1+P_2-P_3-P_4)|M_{12\rightarrow 34}|^2
{sin^2\theta\over 2}.
\end{eqnarray}

\vspace{2cm}

{\large \bf {III. RELAXATION TIME FOR QUARKS AND GLUONS}}

In this section  we use Eq.(6) to calculate the thermal relaxation time 
for quarks and gluons in chemically non-equilibrated plasma.
As thermalization is achieved mainly through
momentum changes in elastic scattering  to the leading order.
Here we only  consider
 the contribution of the elastic scattering.
In the lowest order the relevant matrix elements $M^2$
are [18]
\begin{equation}
M^2_{gg\to gg}={9\over 2}[3-{ut\over s^2}-{us\over s^2}-{st\over u^2}],
\end{equation}
\begin{equation}
M^2_{gq \to gq}=-{4\over 9}{u^2+s^2\over us}+{u^2+s^2\over t^2},
\end{equation}
\begin{equation}
M^2_{q_1q_2\to q_1q_2}={4\over 9}{s^2+u^2\over t^2}.
\end{equation}
Using the definition of the differential cross section[19]
\begin{equation}
{d\sigma\over dt }={g^4\over 16\pi s^2}|M|^2
\end{equation} 
for gluons, we obtain
 \begin{eqnarray}
{1\over \tau_{g}}=&&16\int {d^3k\over (2\pi)^3}f_g^e(k)[1+f_g^e(k)]
\int dt ({d\sigma\over dt })_{gg\to gg}{2tu\over s^2}\nonumber\\
&&+12N_f\int {d^3k\over (2\pi)^3}_qf^e(k)[1-f_q^e(k)]
\int dt ({d\sigma\over dt })_{gq\to gq}{2tu\over s^2},
\end{eqnarray}
Where $f_g$ and $f_q$ are the chemical non-equilibrated
distribution function of gluons and quarks respectively.


The interaction rates diverges at small $k$. As has been discussed,
for  a chemically equilibrated plasma, it can be regulated by the
Debye mass which is the gluon self energy at zero momentum
in the high temperature limit[20].
For a chemically non-equilibrated plasma,  
we use the non-equilibrium Debye mass as the infrared regulator,
\begin{eqnarray}
m_{DNEQ}^2
=4\pi(\lambda_g+{N_f\over 6}\lambda_q)\alpha_s T^2.
\end{eqnarray}

In the following calculation, we assume $f_i$ can be  approximated by
its factorized form[10],
\begin{equation}
f_i=\lambda_i{1\over e^{\beta p_i}{\pm}1}.
\end{equation}
Using Eqs. (11-13) and after some direct 
calculations, we obtain the thermal relaxation time for gluons 
\begin{eqnarray}
{1\over \tau_{g}}=&&{72T^3\over \pi s_{gg}}\alpha_s^2
[2\lambda_g\xi(3)-2\lambda_g^2\xi(3)+\lambda_g^2{\pi^2\over 3}]
[ln(s_{gg}/q^{NEQ})-19/15]\nonumber\\,
&&+{24N_fT^3\over \pi s_{gq}}\alpha_s^2
[{3\over 2}\lambda_q\xi(3)-{3\over 2}\lambda_q^2\xi(3)+\lambda_q^2{\pi^2\over 6}]
[ln(s_{gq}/q^{NEQ})-1.26],
\end{eqnarray}
where $\xi(3)=1.202$,
the infrared cutoff $q^{NEQ}=m_{DNEQ}^2$.
We assume that $s_{gg} and  s_{gq}$ in the above equation
can be  replaced by its thermal average value,
$<s_{gg}>=2<p_g><p_g>$,$<s_{gq}>=2<p_g><p_q>$ where the expression for 
$<p_i>$ is
\begin{equation}
<p_i>={\int {d^3 p\over (2 \pi)^3}p_i f(p_i)\over 
\int {d^3 p\over (2 \pi)^3} f(p_i)}.
\end{equation}
It can be seen from  Eq(15) that $<s>$
is approximately the same for the chemically non-equilibrated
 and equilibrated  gluon gases  
i.e., $<s_{gg}>=14.59T^2$,$<s_{gq}>=16.96T^2$.

For the thermal relaxation time of  quarks we  consider the
processes  $q q\to q q$, $q \bar q \to q\bar q$
and $q g\to q g $.
The thermal relaxation time for quarks we obtained is
\begin{eqnarray}
{1\over \tau_{q}}=&&{32 T^3\over \pi s_{gq}}\alpha_s^2
[2\lambda_g\xi(3)-2\lambda_g^2\xi(3)+\lambda_g^2{\pi^2\over 3}]
[ln(s_{gq}/q^{NEQ})-19/15]\nonumber\\,
&&+{32 N_fT^3\over 3\pi s_{qq}}\alpha_s^2
[{3\over 2}\lambda_q\xi(3)-{3\over 2}\lambda_q^2\xi(3)+\lambda_q^2{\pi^2\over 6}]
[ln(s_{qq}/q^{NEQ})-1.26]
\end{eqnarray}
Where $s_{qq}=19.72 T^2$ in the above equation.

\vspace{2cm}

{\large {\bf IV.  NUMERICAL RESULTS AND DISCUSSIONS}}

The thermal relaxation time depends on the fugacity of gluons and quarks.
To completely determine it we must know the  initial conditions of heavy 
ion collisions, which only can be obtained from models now.
 Focusing on the
effect of the chemical non-equilibration on the thermal 
relaxation time in this paper,
we give a general discussion here.
 Fig. 1-2 show 
the ratio of the thermal relaxation time in chemically nonequilibrium
QGP to the one in chemical equilibrium, $\tau^g_{NEQ}/\tau^g_{EQ}$
and $\tau^q_{NEQ}/\tau^q_{EQ}$,v.s. fugacity $\lambda_g , \lambda_q$
respectively. 
It is found that the results depend weakly on the ratio  between the 
gluon fugacity and quark fugacity. We give the results of two cases,
$\lambda_q=\lambda_g $ and
$\lambda_q={1\over2}\lambda_g$.
$\tau^{NEQ}/\tau^{EQ}$ decreases as the fugacity increases
in both cases.
 There are  corresponding  threshold fugacities $\lambda_g^*$ and $\lambda_q^*$,
For $\lambda_g<\lambda_g^*$ ($\lambda_q<\lambda_q^*$),
$\tau_g^{NEQ}/\tau_g^{EQ}$ ($\tau_q^{NEQ}/\tau_q^{EQ}$)
depends strongly on the fugacity,
while for  $\lambda_g>\lambda_g^*$ ($\lambda_g>\lambda_g^*$), the ratio is almost 1.
These threshold fugacities depend on the QCD coupling constant.
From Fig. 1-2, one can also see that
the bigger the coupling constant, the smaller the threshold fugacity.
The results of Fig.1-2 can be explained as follows:
When the system is approaching  chemical equilibrium, more and more
partons are produced and the fugacity increases continuously. 
The parton density of chemically non-equilibrated plasma is 
smaller than the one in the equilibrium state at the same temperature.
As a result, the relaxation time in  the 
chemically non-equilibrated  plasma is  longer than the one 
in  chemical equilibrium. We also check the
influences of using non-equilibrium Debye mass as the infrared cutoff
and the result is shown Fig. 3.
Using the nonequilibrium Debye mass as the infrared regulator
restrains the increase of $\tau^{NEQ}/\tau^{EQ}$
with decreasing fugacity strongly.

Based on
parton interaction rates calculated for the chemically equilibrated
plasma,
Shuryak argued that equilibration of the plasma proceeds via two stages
in "hot gluon scenario"  [21], where gluons equilibrate much faster than
quarks.
To investigate if the situation is changed for a chemically non-equilibrated
plasma we also calculate the ratio $\tau_q/ \tau _g$
in a chemically non-equilibrated plasma. The results are shown in Fig.4.
We find that the ratios are around 2 and do not change much with the fugacity.
although it changes a little more for longer coupling constants.
From our calculation  we can conclude that there is also  the 
two-stage equilibration 
in a chemically non-equilibrated plasma, i.e., 
gluons also equilibrated much faster
than quarks.

As a final remark, we should point out that only the 
contribution of  the elastic scattering to the thermal relaxation
is taken into account  in this work. 
The contribution of the inelastic scattering should also be 
considered  in the future study.

\section*{Acknowledgement}

This work  is supported in part by the National Nature Science Foundation
of China. We would like to thank Y. Pang , X. Q. Li, A. Tai and 
X. X. Yao for useful discussions.

\newpage

\centerline{\bf  Figure Capure:}

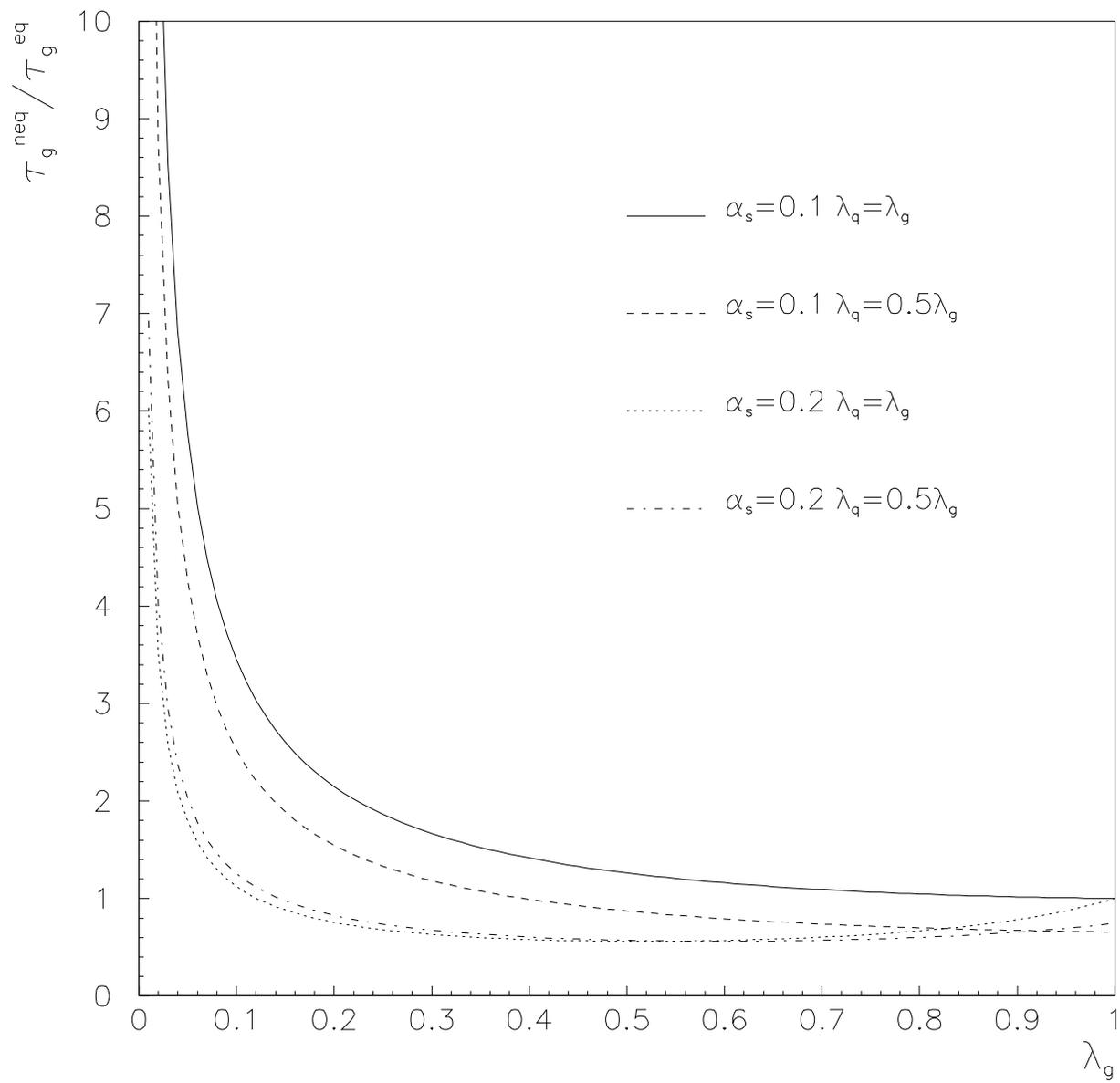
Fig.1.   The ratio of the thermal relaxation time of gluons in chemically 
nonequilibrium plasma  to the one in chemical equilibrium, 
$\tau_g^{NEQ}/\tau_g^{EQ}$, v.s. fugacity $\lambda_g$. 

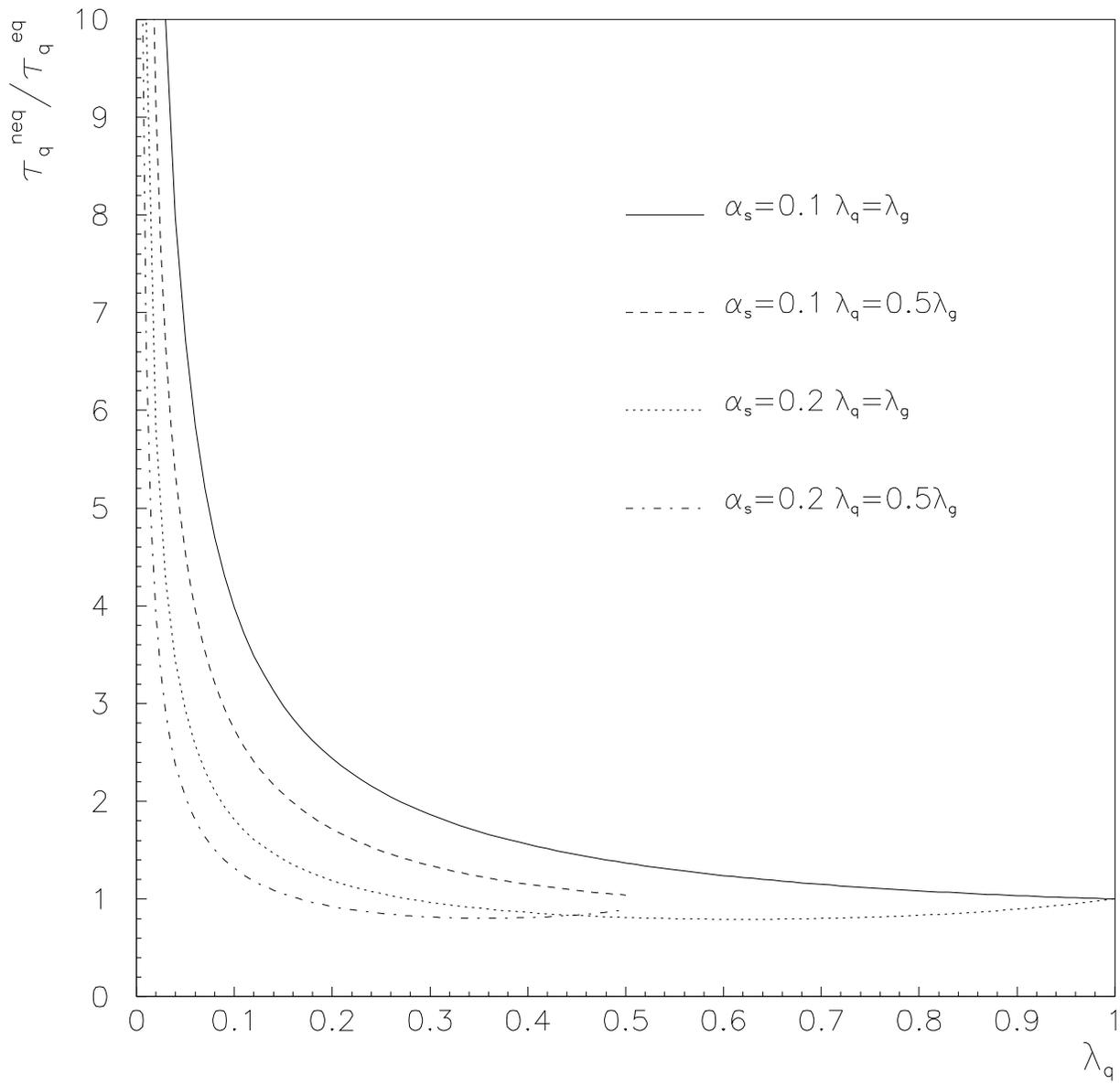
Fig.2.   The ratio of the thermal relaxation time of quarks in chemically 
nonequilibrium plasma  to the one in  chemical equilibrium, 
$\tau_q^{NEQ}/\tau_q^{EQ}$, v.s. fugacity $\lambda_q$.

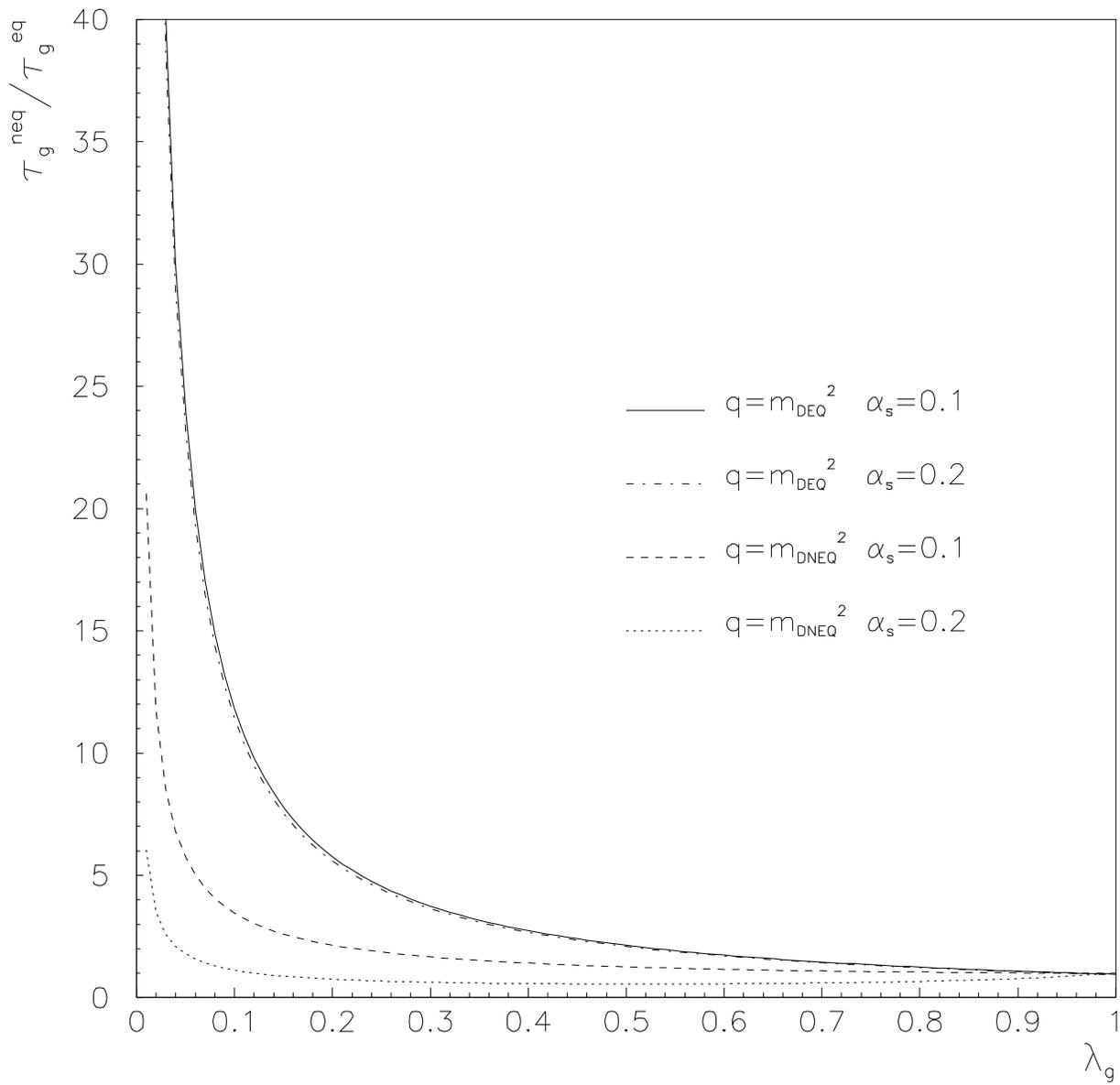
Fig.3. The thermal relaxation time obtained using non-equilibrium 
Debye mass as the infrared regulator comparing  with the one using
equilibrium Debye mass.

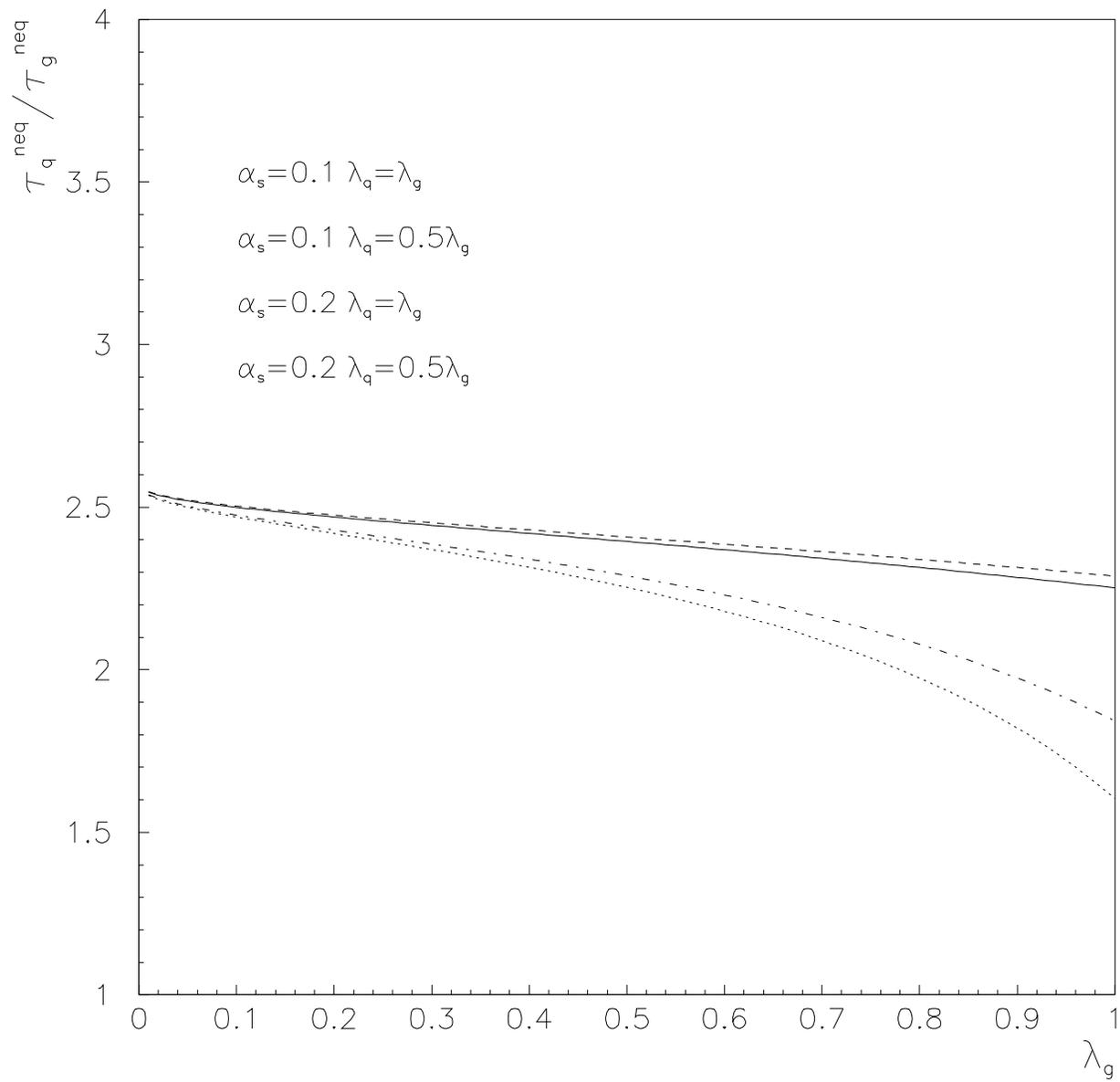
Fig.4. The ratio of the thermal relaxation time of quarks  to
 the thermal relaxation time of gluons   in chemically
nonequilibrium plasma,
$\tau_q^{NEQ}/\tau_g^{NEQ}$, v.s. fugacity $\lambda_g$.


\end{document}